\documentclass[pra,letterpaper,aps,10pt,superscriptaddress,twocolumn,floatfix]{revtex4}
\usepackage{amsmath,graphicx,amssymb,braket,xcolor,subfigure,upgreek}
\usepackage[colorlinks, linkcolor=blue, citecolor=blue, urlcolor=blue, breaklinks=true]{hyperref}

\newcommand{\Sch}{{Schr\"{o}dinger }}

\newcommand{\w}{{\omega}}

\begin{document}

\title{Broadband photon-photon interactions mediated by cold atoms in a photonic crystal fiber}

\author{Marina Litinskaya}
\author{Edoardo Tignone}
\author{Guido Pupillo}
\affiliation{icFRC, IPCMS (UMR 7504) and ISIS (UMR 7006), Universit\'e de Strasbourg and CNRS, 67000 Strasbourg, France.}

\date{\today}

\begin{abstract}
We demonstrate theoretically that photon-photon attraction can be engineered in the continuum of scattering states for pairs of photons propagating in a hollow-core photonic crystal fiber filled with cold atoms. The atoms are regularly spaced in an optical lattice configuration and the photons are resonantly tuned to an internal atomic transition. We show that the hard-core repulsion resulting from saturation of the atomic transitions induces bunching in the photonic component of the collective atom-photon modes (polaritons). Bunching is obtained in a frequency range as large as tens of GHz, and can be controlled by the inter-atomic separation. We provide a fully analytical explanation for this phenomenon by proving that correlations result from a mismatch of the quantization volumes for atomic excitations and photons in the continuum. Even stronger correlations can be observed for in-gap two-polariton bound states. Our theoretical results use parameters relevant for current experiments with Rb atoms excited on the D2-line.

\end{abstract}

\pacs{{\color{red}???????????????}}

\maketitle

\section{Introduction}

There is growing interest in realising strongly interacting photons \cite{Chang,Guerreiro} for applications in quantum information processing \cite{Claudon, Atach, Grafe, Baur, Tiecke, Sprague, Kupchak, Gorniaczyk}, quantum metrology \cite{Chunnilall, Motes, Napolitano} and many-body physics \cite{Carusotto,Chang2008,Otterbach}. Photonic non-linearities are often induced by coupling photons to two-level emitters, as achieved in atomic and molecular setups with single emitters in a cavity configuration~\cite{HarocheRMP,Volz}. Alternatively, photonic non-linearities can result, e.g., from the anharmonicity of the multi-excitation spectra in the Jaynes-Cummings Hamiltonian \cite{Birnbaum, Dayan, Naidon, Reinhard, Devil}, or from dipolar or van-der-Waals interactions between atoms, such as in Rydberg atoms under condition of electromagnetically induced transparency (EIT)~\cite{Adams, Gorshkov, Vuletic, Vuletic1, Weidemuller}. Recent groundbreaking experiments \cite{Vuletic,Vuletic1}, in particular, have demonstrated attraction of photons caused by formation of bound bipolariton states~\cite{Buchler,Maghrebi}. 

In this work we propose the observation of photon-photon interactions and bunching in an ordered ensemble of two-level atoms confined to one-dimension (1D) and resonantly coupled to the transverse photons of a cavity [Fig.~\ref{f-spaceholder1}(a)].
The interaction of atoms and light in the strong exciton-photon coupling regime results in the formation of a doublet 
of polaritonic modes [Fig.~\ref{f-spaceholder1}(b)], corresponding to coherent superpositions of photonic and collective atomic excitations (excitons). In our scheme the non-linearity results solely from the kinematic interaction between these excitons: The latter amounts to a hard-core repulsion and is due to atomic saturability, as one atom can accommodate at most one excitation.

The existence of kinematic interaction in solids has been known for decades \cite{book}, however, was always considered as a very weak effect. In contrast, here we demonstrate that kinematic interaction in a cold atom setup may lead to a pronounced bunching in the photonic component of the coupled polaritonic states. As opposed to narrow bound state resonances in the MHz frequency range typical, e.g., of Rydberg atoms, this bunching appears in the continuum of unbound two-polariton states, and can be observed in a broad GHz frequency range for parameters within the reach of current experimental technologies.Via the exact solution for the subsystem consisting of excitons uncoupled from light, we demonstrate that the bunching is the result of the mismatch of the quantization volumes for states with (excitons) and without (photons) hard core repulsion. Due to the broadband nature of the effect, this type of non-linearity is expected to be comparatively resilient against decoherence. We conclude by discussing the occurrence of bound two-polariton states within spectral gaps.

The scheme we have in mind consists of two-level atoms trapped in a 1D optical lattice in the Mott insulator state (i.e. with one atom per lattice site), and confined inside a 1D resonant cavity. For concreteness, in this work we consider the D2-line of Rb atoms placed into a hollow-core photonic crystal fibre -- as in the experiment \cite{Sr} with Sr. If the cavity losses are low and atoms are well-ordered, in this geometry one can work near the atomic transition without imposing the EIT condition to eliminate absorption: Polaritonic states are coherent superpositions of photonic and collective atomic excitations, as opposed to incoherent absorption of individual atoms.
Besides hollow-core fibers \cite{Perrella, Sprague, Venkataraman,Epple,Na,Blatt,Bajcsy,Sr},
one can think of other implementations of this scheme, as recent theoretical studies have investigated a variety of systems that allow for coupling photons to an ensemble of two-level emitters in a 1D configuration~\cite{Hafezi,Hafezi11,Kimble,Douglas,Zoubi,Pletyukhov,Yu}. 
Solid-state realizations are also possible, e.g., using Si vacancies in photonic crystals \cite{Ahar,Ried}.

\section{Model}
The setup consists of $N$ atoms trapped on a lattice and coupled to a cavity. Its Hamiltonian is




\begin{equation}\label{Hamiltonian}
\begin{split}
H &= E_0 \sum_s P_s^{\dagger} P_s + t \sum_{s} \left(P_s^{\dagger}P_{s+1} + P_s^{\dagger} P_{s-1} \right) \\
&+ \sum\limits_{q_\nu} E_p(q_\nu) b^{\dagger}(q_\nu) b(q_\nu) \\
&+ g \sum\limits_{s,q_\nu}
\left[P_s^{\dagger} b(q_\nu) e^{iqs}+P_s b^{\dagger}(q_\nu)
e^{-iq_\nu s}\right].
\end{split}
\end{equation}

Here, $P_s$,$P_s^{\dagger}$ and $b(q_\nu)$,$b^{\dagger}(q_\nu)$ destroy and create an atomic excitation at site $s$ and a photon with the wave vector $q_\nu$ along the cavity axis, respectively; $E_p(q_\nu) = c\sqrt{q_\nu^2 + q_\perp^2}$ is the photon energy,
$q_\nu = 2\pi\nu/(Na)$ ($\nu$ is an integer, $a$ is the inter-particle spacing, $c$ is the speed of light), and $q_\perp$ is the transverse photon momentum [for the lowest mode of a perfect open cylindrical resonator of radius $R$, it is found from the first zero of
the function $J_0(q_\perp R)$]; $E_0$ is the atomic transition frequency, $t \propto d^2/a^3$ is the hopping energy for the atomic excitations in the nearest neighbor approximation, $g = d \sqrt{2\pi E_0/V}$ is the atom-light coupling constant, with $d$ the transition dipole moment and $V = \pi R^2 N a$ the volume. We note that while the atomic part of equation~\eqref{Hamiltonian} is readily diagonalized by Frenkel exciton operators \cite{book}  $P(q_\nu) = \frac{1}{\sqrt{N}} \sum_n P_n e^{-iq_\nu n}$  describing extended wave functions resulting from exciton hopping, the second term in the Hamiltonian is essentially negligible, since the bare exciton dispersion is much weaker than the polariton dispersion originating from light-matter coupling.\\

In the following, we solve the \Sch equation in the two-particle subspace, where a wave function reads
\begin{equation}\label{Psi}
| \Psi \rangle = \sum_{nm} \left\{ \frac{A_{nm}}{\sqrt{2}}
\ket{b_n b_m} + B_{nm} \ket{b_n P_m} + \frac{C_{nm}}{\sqrt{2}}
\ket{P_n P_m} \right\}.
\end{equation}
Here, $b_n =  \sum_{q_\nu} b(q_\nu) e^{iq_\nu n}/\sqrt{N}$, while $A_{nm}, B_{nm}=B_{nm}^S + B_{nm}^A$ and $C_{nm}$ are the amplitudes for finding two relevant states (photons or atomic excitations) at sites $n,m$ (the superscripts $S,A$ stay for symmetric/antisymmetric; the amplitudes $A$ and $C$ are always symmetric). We recast the \Sch equation as a set of equations for $A, B$ and $C$, and solve it in terms of total and relative wave vectors of two particles, $K_{\nu^{'}}=q_{\nu_1}+q_{\nu_2}$ and $k_{\nu}=(q_{\nu_1}-q_{\nu_2})/2$, respectively, where only $K_{\nu^{'}}$ is a good quantum number. Below we consider in detail the case $K_{\nu^{'}} = 0$, and briefly discuss $K_{\nu^{'}} \neq 0$. For $K_{\nu^{'}} = 0$, $q_{\nu_2} = - q_{\nu_1}$, and $k_\nu = 2\pi\nu/(Na)$ with an integer index $\nu = (-N/2,N/2]$ (a detailed discussion of the quantum numbers describing $K$ and $k$ for bosons on a lattice is provided by Javanainen, J. \textit{et al.}~\cite{K and k}). We obtain:

\begin{equation}\label{Schroedinger-2}
\begin{split}
E_\rho A_\rho(k_\nu) &= 2E_{p}(k_\nu) A_\rho(k_\nu) + G\sqrt{2} B_\rho(k_\nu),\\
E_\rho B_\rho(k_\nu) &= [E_{e}(k_\nu) + E_{p}(k_\nu)] B_\rho(k_\nu)  \\
& +G\sqrt{2} [A_\rho(k_\nu) + C_\rho(k_\nu)],\\
E_\rho C_\rho(k_\nu) &= 2{E_{e}(k_\nu)} C_\rho(k_\nu) + G\sqrt{2} B_\rho(k_\nu) +
S_\rho,\\
\end{split}
\end{equation}
where $E_e(k_\nu) = E_0 + 2t\cos ak_\nu$ is the exciton energy, $\rho$ labels the two-polariton state and
\begin{equation}\label{S(k)}
S_\rho = -\frac{ G\sqrt{2}}{N} \sum\limits_{q_\nu} B_\rho(q_\nu) -
\frac{4t}{N}\sum\limits_{q_\nu} C_\rho(q_\nu) \cos aq_\nu
\end{equation}
is a $k_\nu$-independent term accounting for polariton-polariton scattering due to the kinematic interaction; $B = B^S$ and $B^A = 0$ for $K_{\nu^{'}}=0$; $G = g\sqrt{N}$ is the collective atom-light coupling constant.

\begin{figure}[t]
\centering
\includegraphics[width=\linewidth]{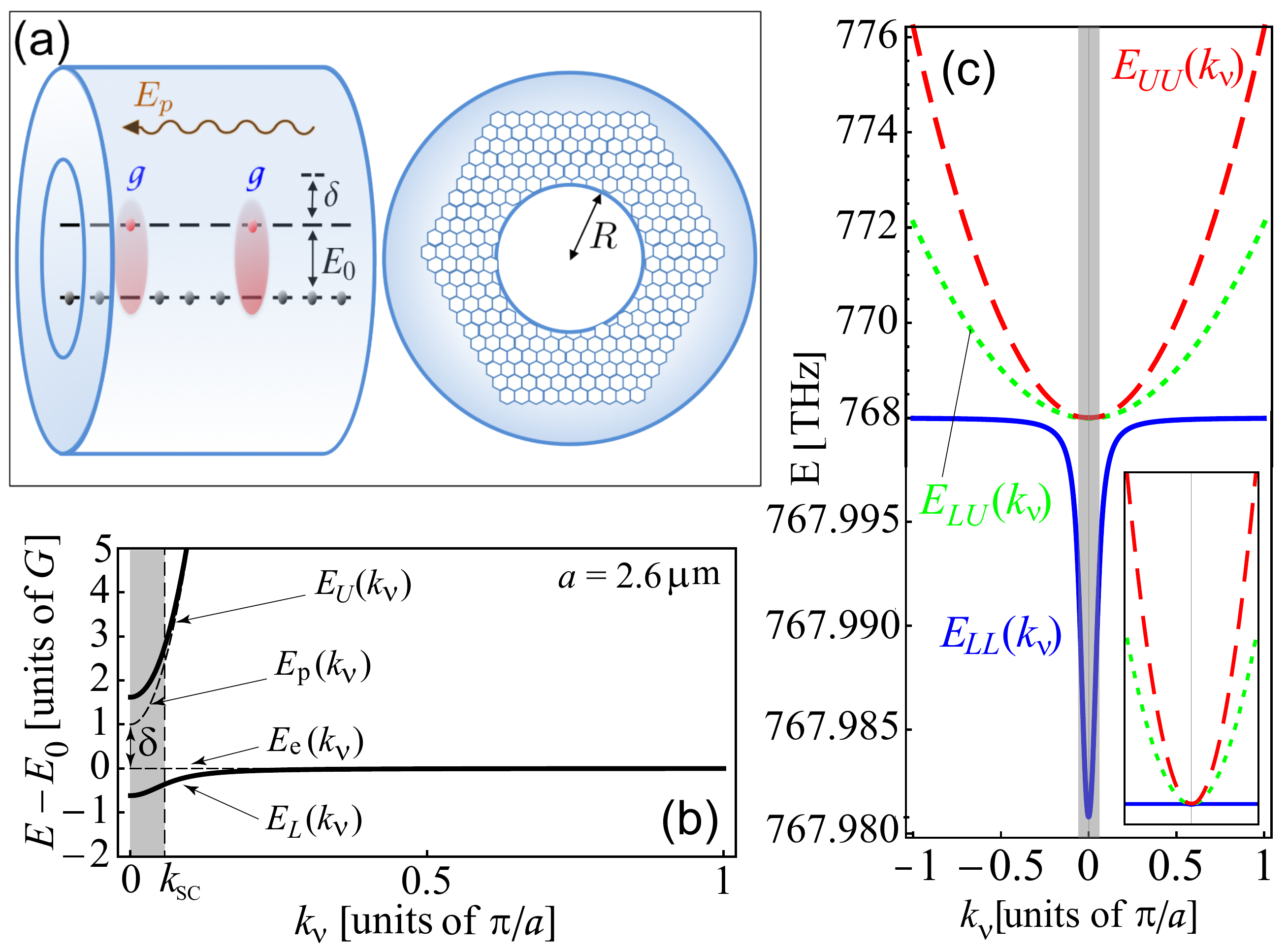}
\caption{({\bf a}) Sketch of the two-level atomic ensemble embedded into a cylindrical cavity showing main parameters (see text). ({\bf b}) Dispersion of lower and upper polaritons, $E_L(k_\nu)$ and $E_U(k_\nu)$, vs.~those of uncoupled exciton and cavity photon, $E_e(k_\nu)$ and $E_p(k_\nu)$, with positive detuning $\delta=E_p(0)-E_0$. Shade marks the strong coupling region approximately restricted by $k_{SC}$. ({\bf c}) Non-interacting two-polariton states: $E_{LL}(k_\nu)=2E_L(k_\nu)$, $E_{LU}(k_\nu)=E_L(k_\nu)+E_U(k_\nu)$, $E_{UU}(k_\nu)=2E_U(k_\nu)$, $a = 2.66~\mu$m, $\delta=0$. Note different energy scales for lower and upper halves of the plot. Inset shows three two-polariton branches plotted together at the same energy scale.}
\label{f-spaceholder1}
\end{figure}

For $S_\rho = 0$, equations~\eqref{Schroedinger-2} describe non-interacting polaritons with dispersion determined by the condition
\begin{equation}\label{Delta}
\begin{split}
&[E - 2E_L(k_\nu)][E - E_L(k_\nu)-E_U(k_\nu)]\\
&\times[E - 2E_U(k_\nu)] \equiv \Delta(E,k_\nu) = 0.\\
\end{split}
\end{equation}
Here, $E_{L}(k_\nu)$ and $E_{U}(k_\nu)$ are the energies of the lower ($L$) and upper ($U$) polaritons, respectively, with
\begin{equation}
E_{L,U}(k_\nu) = \frac{1}{2} \left\{ E_e + E_p \mp \sqrt{(E_e-E_p)^2 +4 G^2} \right\}.
\end{equation}
Figure~\ref{f-spaceholder1}(b) shows that the Brillouin zone can be roughly divided into two distinct regions: The strong-coupling region with $k_\nu < k_{SC}$ near the atom-photon resonance [shaded region in Figs.~\ref{f-spaceholder1}(b) and (c)], and the region with $k_\nu > k_{SC}$, where polaritons essentially behave as uncoupled exciton and photon. The characteristic wave vector $k_{SC} = 2\sqrt{E_0 G}/c\hbar$ is determined by the condition $E_L(k_{SC}) = E_0$ with the parabolic approximation for $E_L(k_{\nu})$.

For finite kinematic interaction $S_\rho \neq 0$ and the solutions of equations~\eqref{Schroedinger-2} are wave packets of free-polariton states. Below we demonstrate that correlations between photons arise as a result of constructive interference among several  components of these wave packets. This effect is more prominent the larger the strong coupling region. In the following, we explain it by first solving analytically the \Sch equation for bare excitons uncoupled from photons, and then by describing the effects of strong coupling of excitons to photons. The latter results in photonic bunching in a broad frequency range in the continuum. The existence of bound two-photon states within polaritonic gaps is discussed  towards the end of the work.\\

\section{Two-photon correlations}

Equations (\ref{Schroedinger-2}) can be solved analytically. By using the equality $\sum_{k_\nu} C(k_\nu) \equiv 0$, which follows from the kinematic interaction constraint $\sum_{k_\nu} C(k_\nu) = C(n=0) \equiv 0$ ($n$ is the relative distance between two excitations in the site representation), we reduce the problem to three independent equations of the form
\begin{equation}
x_\rho(k_\nu) = \frac{1}{N}\sum_{k_{\nu^{'}}} x_{\rho}(k_{\nu^{'}}),
\end{equation}
with $x_\rho(k_\nu)$ representing the quantities $A_\rho(k_\nu) \Delta(E_\rho,k_\nu)$,
$B_\rho(k_\nu)$$\Delta(E_\rho,k_\nu)/[E_\rho-2E_p(k_\nu)]$ and $C_\rho(k_\nu) \Delta(E_\rho,k_\nu)/\phi(E_\rho,k_\nu)$,
where $\Delta(E_\rho,k_\nu)$ is defined in equation (\ref{Delta}), and
$\phi(E,k_\nu) = [E - 2E_p(k_\nu)][E-E_p(k_\nu)-E_e(k_\nu)] - 2G^2$.
Then each equation is solved by $x_\rho(k_\nu) = const(\rho)$. Introducing the normalization constant
\begin{equation}
c_\rho = \left(\sum_{k_\nu}\frac{ [\phi^2(E_\rho,k_\nu) + 2G^2(E_\rho - 2E_p)^2 + 4G^4]}{\Delta^2(E_\rho,k_\nu)}\right)^{-1/2},
\end{equation}
we finally write

\begin{equation}\label{ABC-exact}
\begin{split}
A_\rho(k_\nu) &= \frac{2G^2 c_\rho}{\Delta(E_\rho,k_\nu)},\\
B_\rho(k_\nu) &= \frac{[G\sqrt{2}(E_\rho - 2E_p(k_\nu))c_\rho]}{\Delta(E_\rho,k_\nu)},\\
C_\rho(k_\nu) &= \frac{\phi(E_\rho k_\nu)c_\rho}{\Delta(E_\rho,k_\nu)}.\\
\end{split}
\end{equation}

\begin{figure}[t]
\centering
\includegraphics[width=\linewidth]{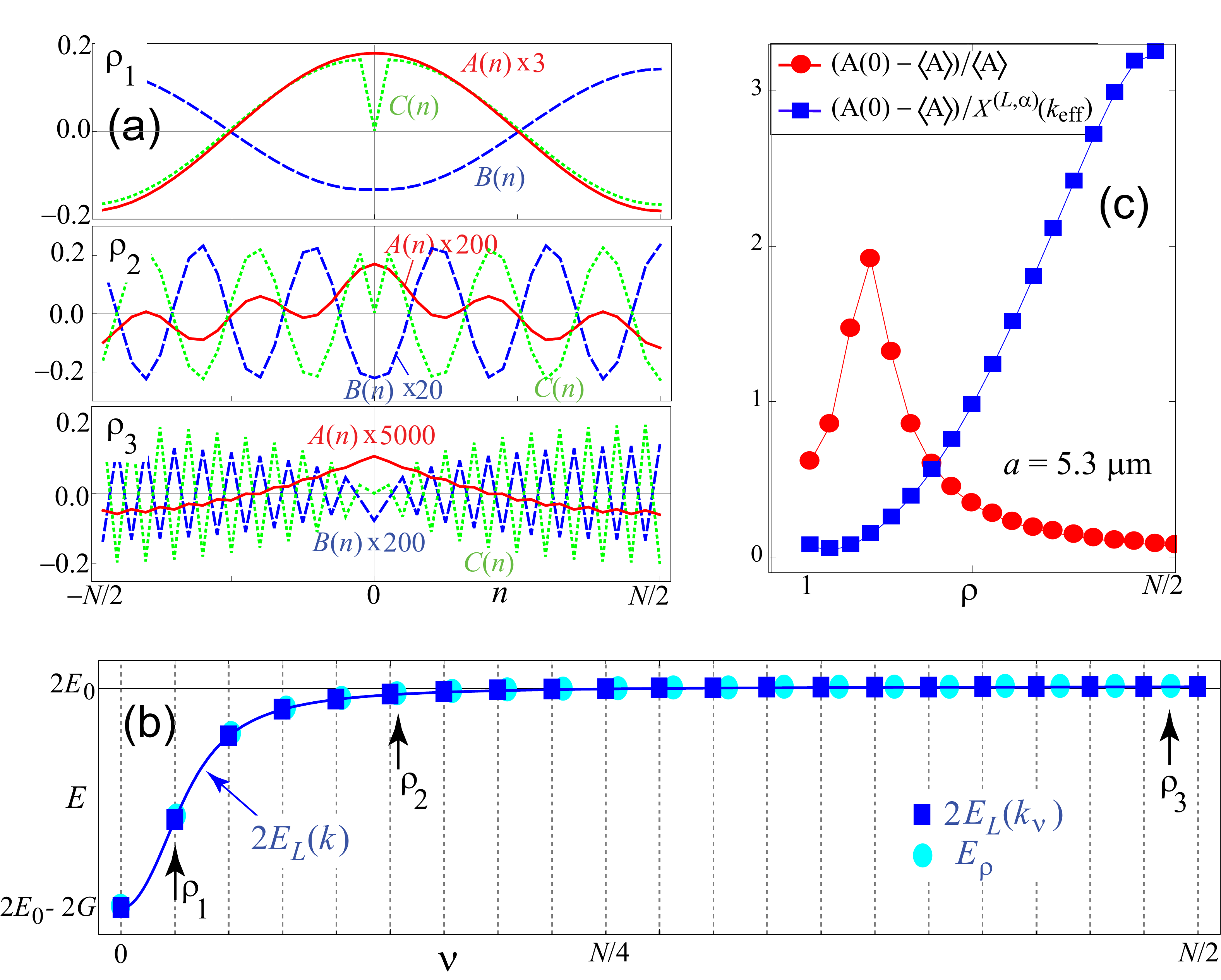}
\caption{({\bf a}) Amplitudes $X(n)-\langle X(n) \rangle$, $X = A$ (red), $B$ (blue dashed) and $C$ (green dotted), scaled by the factors shown in each figure, for three two-polariton states marked by arrows in panel (b); $n$ is the relative distance, $N=40$, $a = 5.3~\mu$m. ({\bf b}) Exact two-polariton energies (cyan), compared to the energies of non-interacting polaritons $2E_L(k_\nu)$ (blue squares). Cyan points are plotted at the positions given by $k_{\rm eff}(\rho)$. ({\bf c}) Bunching strength as function of state number, absolute value (red circles) and scaled with the account of photonic amplitude of the state (blue squares).}
\label{f-spaceholder2}
\end{figure}

Figure~\ref{f-spaceholder2}(a) shows example results for the real-space Fourier transform of the amplitudes \eqref{ABC-exact} for the
three states from the bottom, middle and top of the two-polariton lower-lower- (LL-) band, whose energies  are indicated by arrows in Fig.~\ref{f-spaceholder2}(b). For small $\rho$ (upper plot) all amplitudes resemble free wave states with a sharp dip in the two-exciton amplitude $C(n)$ at $n=0$,
which is a result of the hard-core constraint. With the increase of $\rho$ within the LL-band, the amplitudes $B(n)$ and $C(n)$
demonstrate modulated oscillations, with the excitation probability increasing towards larger separations. In contrast, the two-photon amplitude $A(n)$ stops oscillating for larger $\rho$ and displays a peak-like feature centered at $n=0$ (see lower plot): The two-photon amplitude thus demonstrates bunching in the presence of repulsive kinematic
interaction among atoms.\\

In order to clarify the behavior of the amplitudes, we first solve
the \Sch equation
\begin{equation}
\begin{split}
E_\mu C^{(ex)}_\mu(n) &= (1 - \delta_{n0})
\Big\{ 2E_0 C^{(ex)}_\mu(n) + 2t\\
&\times\left[ C^{(ex)}_\mu(n+1) +
C^{(ex)}_\mu(n-1)\right] \Big\}
\end{split}
\end{equation}
for two bare excitons interacting via the
kinematic interaction in the nearest neighbor approximation (i.e., in the absence of coupling to photons). Remarkably, we
find that this latter equation is analytically solvable (see Appendix A for details).
Due to the exclusion of the state $C^{(ex)}(n=0)$  caused by the kinematic interaction, the states are now described by a new set of wave vectors $\{\kappa_\mu\}$, with
\begin{equation}\label{kappa_rho}
\kappa_\mu = \frac{2\pi\mu}{Na},
\end{equation}
and $\mu = [-\frac{(N-1)}{2},
\frac{(N-1)}{2}]$ {\it half-integer}. The set $\{\kappa_\mu\}$ has elements that lie exactly between those of the original set $\{k_\nu\}$. This suggests that the exciton-exciton kinematic interaction, however weak, is a
non-perturbative effect. The new two-exciton eigenenergies are then
$E^{(ex)}_\mu = 2E_0 + 4t \cos a\kappa_\mu$, and their amplitudes read
\begin{equation}\label{Cex(mu)}
C_\mu^{(ex)}(n) = (1-\delta_{n0})\sin a\kappa_\mu |n|.
%
\end{equation}

For finite exciton-photon coupling and $E_\rho \lesssim 2E_L(k_{SC})$, the exciton-photon coupling prevails over exciton-exciton interactions. In this regime each eigenstate is to a good approximation described by a single $k_\nu$, and the $C$-amplitudes behave approximately as $C_\nu(n) \propto (1-\delta_{n0}) \cos ank_\nu$, i.e. as the symmetric part of a plane wave with $2\nu$ nodes. For $E_\rho \gtrsim 2E_L(k_{SC})$, however, the exciton-exciton interaction prevails over light-matter coupling, and the amplitudes $C$ reach the exciton-like limit, where they are well approximated by equation~\eqref{Cex(mu)}. In other words, with the increase of $\rho$, the two-polariton quantum number $\rho$ makes a smooth transition from $\nu$-numbers to $\mu$-numbers.
We find that in all parameter regimes the two-polariton energy is well approximated by the analytical expression $E_\rho = 2 E_L(k_{\rm eff}(\rho))
$,
where
\begin{equation}
k_{\rm eff}(\rho) = \frac{2\pi\rho_*}{Na}, \hskip 0.5cm \rho_* = (\rho -
1)\frac{N/2 - 1/2}{N/2 - 1},
\end{equation}
correspond to an effective wave vector $k_{\rm eff}$ and its label $\rho_*$, respectively, interpolating between the two sets above.
This is shown in Fig.~\ref{f-spaceholder2}(b), where the exact numerical results for the energies from equations~\eqref{Schroedinger-2} (cyan dots) plotted as a function of the effective wave vectors $k_{\rm eff}$ perfectly match the analytical estimates $E_{LL}(k_\nu \to k_{\rm eff}(\rho))$.


In the basis of the original wave vector set $\{k_\nu\}$, the gradual shift to the set $\{\kappa_\mu\}$ corresponds to the formation of wave packets. 
The resonant character of the factor $1/\Delta(E_\rho, k_\nu)$ in equation~\eqref{ABC-exact} implies that $A_\rho(k_\nu)$ is peaked at $k_\nu \sim \pm k_{\rm eff}(\rho)$. However, these components dominate the shape of $A_\rho(k_\nu)$ only for states within the strong coupling region, i.e. when $k_{\rm eff}(\rho) < k_{SC}$. In contrast, for $k_{\rm eff}(\rho) > k_{SC}$, the components $A_\rho(k_\nu \sim k_{\rm eff}(\rho))$ are strongly suppressed, as $1/\Delta(E_\rho, k_\nu) \propto 1/E_U^2(k_\nu) \approx 1/E_{p}^2(k_\nu)$ decays fast outside of the strong coupling region [see inset in Fig.~\ref{f-spaceholder1}(c)]. Therefore, for larger $\rho$ only low-$k_\nu$ states (with $k_\nu \lesssim k_{SC}$) are found to contribute to $A_\rho(k_\nu)$; in other words, higher-$k_\nu$ states are too off-resonant to participate in the formation of the wave packets and, as a result, at the large-$\rho$ amplitudes $A(k)$ develop a single maximum around $k_\nu = 0$. This cusp-like structure of $A_\rho(k_\nu)$ results in the cusp-like shape of $A_\rho(n)$ in real space [Fig.~\ref{f-spaceholder2}(a), middle and lower panels]. This explains the central result of this paper: The mismatch between the quantum numbers describing the interacting ($C$) and non-interacting ($A-B$) subsystems leads to systematic two-photon bunching at $n = 0$. As $A_\rho(n)$ takes its maximum value at the same
separation $n = 0$ for all $\rho$ states with $k_{\rm eff}(\rho)\gtrsim k_{SC}$, we expect that it should not be averaged out by a finite width of the exciting source. In the Appendix B we quantify these arguments and interpret the effect in terms of interference between different $A_\rho(k_\nu)$-components.\\

\section{Controlling two-photon correlations}

Correlations in the continuum of two-polariton states should be observable with current experimental technologies with Rb or Sr atoms in a Mott insulator state with unit filling, placed in a hollow-core fiber, e.g., in the configuration of Okaba \textit{et al.}~\cite{Sr}. For example, choosing the radius of the fiber $R = 0.299~\mu$m, we bring the lowest cavity mode in resonance with the D2 transition in Rb (transition dipole $d = 4.22$~a.u.) at $E_0 = 384$~THz. We quantify the two-photon bunching by the figure of merit $\Delta A_\rho$ defined as
\begin{equation}\label{DeltaA}
\Delta A_\rho = \frac{|A_\rho(n=0)| - \braket{A_\rho}}{\braket{A_\rho}}, \hskip 0.5cm \braket{A_\rho(n)} = \frac{1}{N}\sum_n |A_\rho(n)|
\end{equation}
if the difference is positive, and zero otherwise and plot it in Fig.~\ref{f-spaceholder2}(c) (red circles). The apparent decrease of $\Delta A$ for large $\rho$ is a result of the overall decrease of the photonic wave function in the polaritonic state. To demonstrate that the bunching effect in fact {\it increases} with increasing $\rho$, in Fig.~\ref{f-spaceholder2}(c) we plot  $\Delta A_\rho/ X^{(L,\alpha)}(k_{\rm eff}(\rho))$ (blue squares), where $X^{(L,\alpha)}(k_{\rm eff}(\rho))$ 
is the two-photon component in the two-polariton wave function in the absence of kinematic interaction, as defined in the Appendix B.

Figure~\ref{f-spaceholder3}(a) shows that $\Delta A(E_\rho)$ changes by varying the lattice constant $a$ from 532~nm to 5.32~$\mu$m. Counterintuitively, $\Delta A$ is found to {\it increase} for larger $a$, corresponding to lower atomic density. This is explained by noting that only states within the strong coupling region, which have both finite two-exciton amplitude (to interact effectively) and finite two-photon amplitude (the observable), contribute to $A_\rho(n=0)$. Therefore, photon bunching is most pronounced when $k_{SC}$ is comparable to the size of the Brillouin zone, which can be easily achieved in cold atom experiments. The latter condition
requires larger inter-atomic separations, as $ak_{SC}/\pi \propto a^{3/4}$.  
This also explains why this type of bunching of continuum states cannot be observed in solids: For electronic transitions of $\sim 2$~eV and typical lattice constants $a \sim 5$~\AA, the relative size of the strong coupling region $a k_{SC}/\pi \sim 2 a \sqrt{\w_0 G}/\pi c\hbar \sim 10^{-4}$. On the contrary, Fig.~\ref{f-spaceholder3}(a) shows that in cold atom systems considerable bunching can be realised  in a continuous band of the order of several GHz.

In Figure~\ref{f-spaceholder3}(b) we examine $\Delta A$ vs energy for a few values of the detuning $\delta = E_p(0)-E_0$ between the cavity mode $E_p(0)$ and the excitonic resonance $E_0$, by varying $\delta$ between $-G/2$ and $G$. We find that large negative values of $\delta$ result in wider bunching frequency intervals due to wider $E_{LL}$ bands.

Finally, it is of crucial importance that the bunching survives for states with $K_{\nu^{'}} \neq 0$. The latter correspond to propagation of the center of mass of the two polaritons, and can be directly observed in experiments. The equations for the amplitudes at $K_{\nu^{'}} \neq 0$ are bulky and will be published elsewhere. Here we only demonstrate the existence of bunching in a wide range of $(k_{\nu_1}, k_{\nu_2})$-pairs by plotting in Fig.~\ref{f-spaceholder3}(c) $\Delta A$ as a function of $K_{\nu^{'}}$ for two states, selected so that at $K_\nu = 0$ they correspond to the full squares marked as $\rho_1$ and $\rho_2$ in Fig.~\ref{f-spaceholder3}(a).

\begin{figure}[t]
\centering
\includegraphics[width=\linewidth]{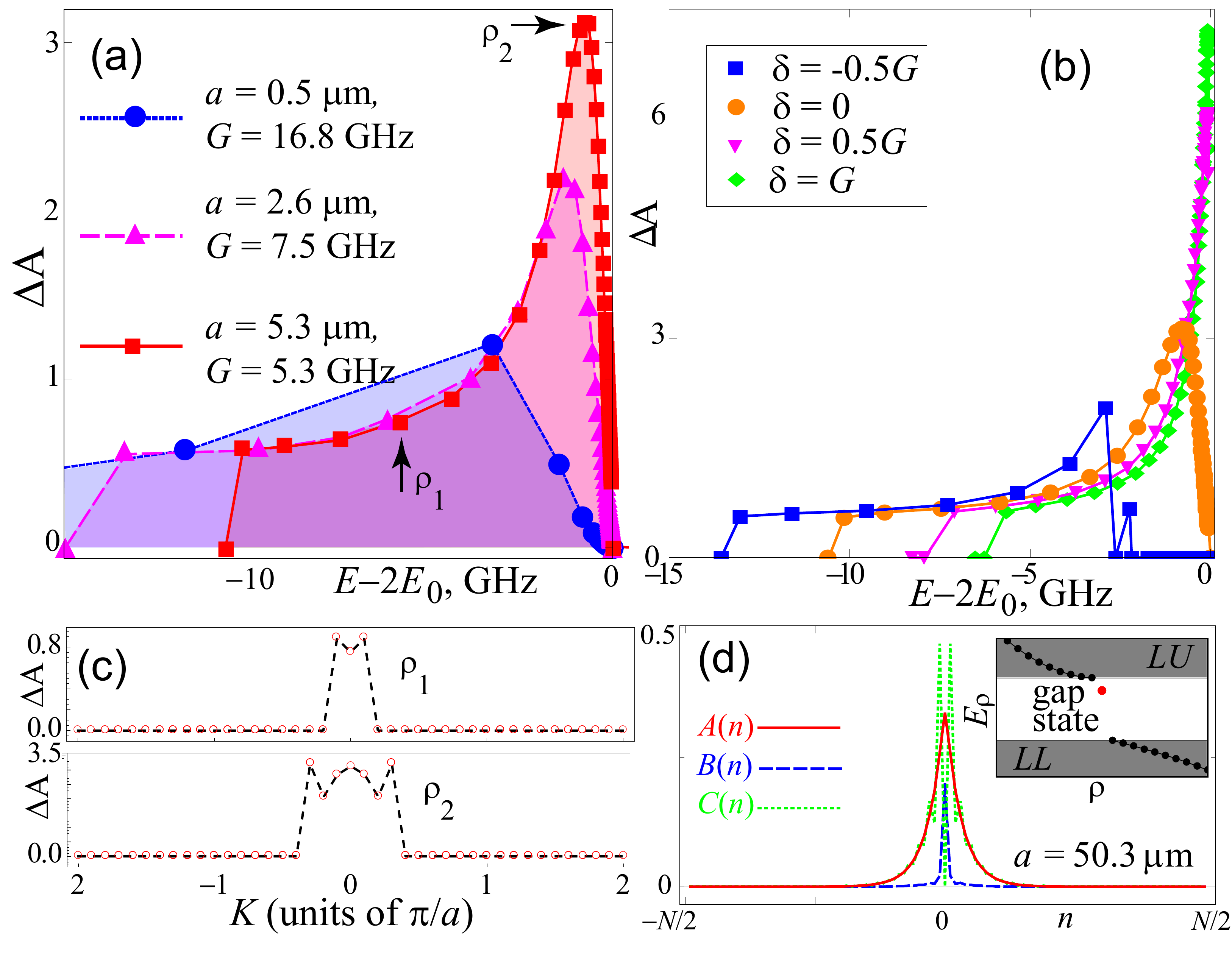}
\caption{({\bf a}) Figure of merit $|\Delta A (n=0)|$ as a function of the state energy for Rb atoms, D2 line, for $a = 532$~nm,
2.66~$\mu$m and 5.32~$\mu$m. ({\bf b})  $|\Delta A (n=0)|$ in the LL-band for $a$ = 5.3~$\mu$m for different values of detuning. {\bf(c)}
$|\Delta A (n=0)|$ as function of $K$ for two states, which at $K_{\nu^{'}} = 0$ correspond to the states marked by $\rho_1$ and $\rho_2$ in
panel (a); $a$ = 5.3~$\mu$m. ({\bf d}) Amplitudes $A(n)$, $B(n)$ and $C(n)$ for the bipolariton gap state; $a$ = 50.3~$\mu$m. Inset displays the
energies $E_\rho$ as a function of the state number $\rho$, with the gap state shown in red.}
\label{f-spaceholder3}
\end{figure}

\section{Gap states}

A gap in the polaritonic spectrum appears naturally when the detuning is positive. In contrast, for $\delta = 0$ and small $a$ the two-polariton spectrum is non-gapped, as the energy for two lower polaritons $E_{LL}(\pi/a)$ equals $E_{LL}(\pi/a)= E_{LU}(0) = 2E_0$, with $E_{LU}$ the energy of the lower-upper- (LU-) band. However if $a$ is so large that $k_{SC} \sim \pi/a$,
then $E_{LL}(\pi/a)< 2E_0$ as the lower polariton does not reach the excitonic limit, and a small gap $\Delta_{LU} = E_{LU}(k_\nu=0) - E_{LL}(k_\nu=\pi/a)$ opens between the LL- and LU- bands
at zero detuning. In this regime one can observe the formation of bunched states also {\it within the gapped region} at the very bottom of the LU-band. Contrary to the discussion above, these are {\it polariton-polariton bound states} with very distinct wave functions, with $A$- and $B$-amplitudes peaked around $n=0$ (the two-exciton $C$-component vanishes at $n=0$ and therefore is maximal for the separation $|n|=1$, in accordance with the hard core restriction) [see Fig.~\ref{f-spaceholder3}(d)]. For moderate $a$, the bound state merges with the LU-band. With the increase of the lattice constant, at $a \sim 25~\mu$m, the bound state splits from the band and shifts into the gap. Further increase of $a$ is accompanied by deeper penetration into the gap, dramatic increase of the photonic bunching in this particular state, and suppression of the bunching in the continuum.

The state above is a gap bipolariton forming under repulsive kinematic interaction. This is similar to the kinematic biexciton appearing in organic crystals with two molecules in a unit cell \cite{KinemBiexc}. However, there the kinematic biexciton overlaps with the continuum band, and can be easily destroyed by, e.g., disorder or coupling to phonons. In contrast, the kinematic bipolariton described here is located in the gap and is thus stable against decoherence. Other types of bound states, both below the LL-continuum and in the polariton gap, can form if the atoms interact via, e.g., dipole forces. This would be analogous to, e.g., the gap bound states found in atomic systems~\cite{GapStates,GapStates2} and Jaynes-Cummings-type models with repulsive interactions~\cite{Wong2011}.

For $a$ large enough the interaction of excitons with photons from higher Brillouin zones becomes possible: An exciton with a wave vector $k_\nu$ is coupled not only to a photon with the same $k_\nu$ but also to photons with $k_\nu \pm 2\pi/a,$ $k_\nu \pm 4 \pi/a$, etc. For the setup described here, this would occur for a large value $a \gtrsim 50~\mu$m. What happens in this regime will be the subject of a further investigation.

\section{Discussion}

In this article we have characterised the correlations that are generated for pairs of 1D photons propagating in a hollow-core crystal fiber and coupled to an ordered atomic ensemble. The correlations resemble two-photon attraction for states in the continuum of scattering states. This two-photon bunching can be observed in GHz frequency window, which greatly exceeds MHz frequency intervals typical for the correlations induced by, e.g., three-level cold Rydberg atoms in the EIT regime, and can be controlled by tuning the inter-atomic spacing. Our results are valid as long as the polaritonic splitting
$2G$ exceeds strongly the sum of the excitonic and photonic broadenings, which can be achieved by using a high-quality cavity and by
careful preparation of the atomic Mott insulator state. Other realizations of this scheme are also possible, as long as  broadenings can be kept smaller than the collective exciton-photon coupling. In addition to 1D hollow-core crystal
fibers, promising candidates are metallic nanowires \cite{Akimov} and nanophotonic waveguides \cite{Goban,Nayak,Vetsch,Goban2012,
Rauschenbeutel,Mitsch,Kien,Chang2008}.

The same effect as demonstrated here, and even in an exaggerated form, can exist for chains of two-level Rydberg atoms coupled to 1D cavity photons. We find that the formation of a large-radius Rydberg blockade sphere considerably enhances bunching, as the effect of the quantization volumes mismatch presented in our work becomes more pronounced. In addition, the long-range dynamical (dipole-dipole or van der Waals) interaction is found to enrich the arising of non-linear effects. These results will be subject of
a forthcoming publication.

\section{Acknowledgements}
We acknowledge support  by  the  ERC-St  Grant  ColdSIM  (No.   307688),  UdS  via  Labex  NIE  and
IdEX, RYSQ.

\appendix

\section{Kinematic interaction for bare excitons}
The \Sch equation for two bare excitons ($G \equiv 0$) on sites $n_1$  and $n_2$ interacting via kinematic interaction is
\begin{equation}\label{eq:Cbareex}
\begin{split}
E C_{n_1n_2}^{(ex)} &= 2E_0 C_{n_1n_2}^{(ex)} + (1- \delta_{n_1n_2})\\
&\times\sum\limits_s \left( t_{n_1s} C_{sn_2}^{(ex)} + t_{n_2s} C_{n_1s}^{(ex)} \right)
\end{split}
\end{equation}
with $t_{ij}$ the long-range hopping energy, and the same-site amplitude chosen as $C^{(ex)}_{nn} = 0$ \cite{Vektaris}. Let $n$ be $n = |n_1-n_2|$, and the index $\mu$ enumerate the eigenstates of this equation. We rewrite equation~\eqref{eq:Cbareex} in the nearest neighbour approximation:
\begin{equation}\label{kinematic_ex}
\begin{split}
E^{(ex)}_\mu C^{(ex)}_\mu(n) &= (1 - \delta_{n0}) \Big\{ 2E_0 C^{(ex)}_\mu(n) +
2t\\
&\times[ C^{(ex)}_\mu(n+1) + C^{(ex)}_\mu(n-1)] \Big\}.
\end{split}
\end{equation}

One can verify that the normalized amplitudes which satisfy equation~\eqref{kinematic_ex} are \begin{equation}\label{Cex(n)}
C_\mu^{(ex)}(n) \equiv  g_n(\mu) =\frac{\sqrt{2}(1 - \delta_{n0})}{\sqrt{N}}\ \sin |n|\kappa_\mu
\end{equation}
with wave vectors $\kappa_\mu$ introduced in equation~(10) in the main text. The basis functions $g_n(\mu)$ form an orthonormal set in both spaces, with the orthonormality conditions reflecting the permutation symmetry and the hard-core condition:
\begin{equation}
\begin{array}{c}
\label{eq:gprops}
\sum_n g_n(\mu_1)g_n(\mu_2)=\delta_{|\mu_1|,|\mu_2|},\\

\\

\sum_\mu {g_{n_1}}(\mu){g_{n_2}}(\mu)=(1-\delta_{n_1 0})\delta_{|n_1|,|n_2|}.\\
\end{array}
\end{equation}

The eigenenergies of two-exciton states are also described by the wave vectors $\kappa_\mu$ as
\begin{equation}\label{Eex}
E^{(ex)}_\mu = 2E_0 + 4t \ \cos a \kappa_\mu.
\end{equation}
These new wave vectors $\kappa_\mu = 2\pi\mu/(Na)$ have a half-integer state index $\mu = [-(N-1)/2, (N-1)/2]$ and lie exactly between the positions of the standard wave vectors $k_\nu$ for non-interacting excitons. As a consequence, the amplitudes $C_\mu^{(ex)}(k_\nu)$ do not have poles, but rather an enhanced components of $k_\nu \approx \kappa_\mu$, as can be seen from the Fourier transform of equation~(\ref{Cex(n)}):
\begin{equation}\label{Cex(q)}
C_\mu^{(ex)}(k) = \frac{\sin a\kappa_\mu \cos (ak_\nu) + (-1)^\mu \sin ak_\nu \sin (ak_\nu N/2)}{\cos ak_\nu - \cos a\kappa_\mu}.
\end{equation}
We conclude that the kinematic interaction is a weak, but absolutely non-perturbative effect for excitons. Let us now discuss the effect of the kinematic interaction for polaritons.

\section{Creation of the wave packets}

Two-polariton states in the presence of kinematic interaction can be viewed as composed of two subsystems: the non-interacting subsystem (consisting of photon-photon and photon-exciton states) is described by the quantum numbers $k_\nu = 2\pi\nu/(Na)$ with integer $\nu$, whereas the interacting subsystem (consisting of exciton-exciton states) is described by $\kappa_\mu = 2\pi\mu/(Na)$ with half-integer $\mu$. In the following discussion we shall stress the role of the two wave vector sets, which will be reflected in the adopted notations. The coupling between these two subsystems is responsible for intermixing the corresponding wave vector sets $\{k_\nu\}$ and $\{\kappa_\mu\}$ and eventually leads to the creation of the wave packets in the ``original" wave vector set $\{k_\nu\}$. In particular, at the lowest energies the coupling of excitons to photons dominates over exciton-exciton interaction, and the corresponding polaritons are better described by $k_\nu$. With the increase of the state number, instead, polaritons enter the exciton-like regime and are better described by $\kappa_\mu$.

We introduce the operator $\alpha_n^\dag$, $\beta_n^\dag$ and $\gamma_n^\dag$, which describe, respectively, creation of two photons, one photon and one exciton, and two excitons separated by a distance $n$. The two particle wave function takes the form $\ket{\Psi}=\sum_s\left[A(s)\ket{\alpha_s}+B(s)\ket{\beta_s}+C(s)\ket{\gamma_s}\right]$, and the Hamiltonian $\tilde H_{\rm eff}=\tilde H_{AB}+\tilde H^{(KI)}_{C}+\tilde H^{(KI)}_{AB-C}$ is made up of three terms:

\begin{equation}\label{H-nu-mu}
\begin{split}
\tilde H_{AB} &= \sum\limits_{n,m}\Big[2E_p(n-m)\alpha_n^\dagger\alpha_m\\
&+\left( E_p(n-m) + E_e(n-m)\right)\beta_n^\dagger \beta_m\Big]\\
& +G\sqrt{2}\sum_n \left[\alpha_n^\dagger \beta_n+\beta_n^\dagger \alpha_n\right],\\
\tilde H^{(KI)}_{C} &=  \sum\limits_{n,m} (1-\delta_{n0})2E_e(n-m)\gamma_n^\dagger\gamma_m,\\
\tilde H^{(KI)}_{AB-C} &= G\sqrt{2}\sum\limits_n (1-\delta_{n0})\left[\gamma_n^\dagger \beta_n+\beta_n^\dagger \gamma_n\right],\\
\end{split}
\end{equation}
where the last one describes the coupling between the interacting ($C$) and non-interacting ($AB$) subsystems. The resulting \Sch equation is identical to the Fourier transform of equations~(3) in the main text.

The first term $\tilde{H}_{AB}$ describes the subspace ``photon-photon $\bigcup$ photon-exciton", and is diagonalized by the operators ${\xi_{\nu}^{(i)}}^\dagger = X^{(i,\alpha)}_{\nu} \alpha_\nu^\dagger + X^{(i,\beta)}_{\nu} \beta_\nu^\dagger$: 
\begin{equation}
\tilde H_{AB}= \sum\limits_{i=L,U} \sum\limits_{\nu} E^{(p,i)}_\nu {\xi^{(i)}_\nu}^\dagger \xi^{(i)}_\nu.
\end{equation}
Here and below $i=(U,L)$ is the index of the polaritonic branch, $\nu = (-N/2,N/2]$ is the free-state wave index, and
\begin{equation}\label{AB-polaritons}
E^{(p,i)}_\nu = E^{(p)}_\nu+\frac{E^{(e)}_\nu + E^{(p)}_\nu \pm \sqrt{\left( E^{(p)}_\nu-E^{(e)}_\nu \right)^2 + 8 G^2} }{2}
\end{equation}
with $E^{(p)}_\nu \equiv E_p(k_\nu)$,  $E^{(e)}_\nu \equiv E_e(k_\nu)$. The energies $E^{(p,i = \{ L,U \})}_\nu$ (with $i=L$ corresponding to ``$-$", and $i=U$ to ``$+$" in the right-hand side) are constructed as sums of energies of one photon and one exciton-polariton with the coupling constant $\sqrt{2}G$, taken at the same wave vector $k_\nu$. They naturally appear as solutions of the first two lines of equations~(3) in the main text with $C \equiv 0$. The photon-photon and photon-exciton amplitudes are
\begin{equation}\label{X}
\begin{split}
&\displaystyle X^{(i,\alpha)}_{\nu} = \sqrt{\frac{\left(E^{(p,i)}_\nu - E^{(p)}_\nu - E^{(e)}_\nu\right)^2}{2G^2 + \left(E^{(p,i)}_\nu - E^{(p)}_\nu - E^{(e)}_\nu \right)^2}},\\
&X^{(i,\beta)}_{\nu} = \sqrt{1 - \left( X_\nu^{(i,\alpha)} \right)^2}.
\end{split}
\end{equation}

The two-exciton part $\tilde H^{(KI)}_{C}$ of the Hamiltonian is instead diagonalized as
\begin{equation}
\tilde H^{(KI)}_{C} =  \sum\limits_{\mu} E_{\mu}^{(ex)} \chi^\dagger_\mu\chi_\mu,
\end{equation}
with energy $E_\mu^{(ex)}$ defined in equation \eqref{Eex}, and
\begin{equation}
\chi_\mu^\dag = \sum\limits_{s = -N/2+1}^{N/2} g_s(\mu)\gamma_s^\dagger.
\end{equation}

Eventually, the interaction Hamiltonian $\hat{H}_{AB-C}$ can be rewritten in terms of $\xi$- and $\chi$-operators as
\begin{equation}
\tilde H^{(KI)}_{AB-C} = \frac{G}{N}\sum\limits_{i = L,U} \sum\limits_{\nu\mu} \Lambda_{\nu\mu}X^{(i,\beta)}_{\nu}\left(\chi_\mu^\dagger \xi^{(i)}_{\nu} + {\xi_{\nu}^{(i)}}^\dagger \chi_\mu\right)
\end{equation}
with coupling coefficients $\Lambda_{\nu\mu}$ given by
\begin{equation}\label{Lambda}
\Lambda_{\nu\mu} = \frac{1}{2}\left[\cot \frac{\pi(\nu+|\mu|)}{N}  - \cot \frac{\pi(\nu-|\mu|)}{N}\right].
\end{equation}
The coefficients $\Lambda_{\nu\mu}$ intermix the wave vector sets $k_\nu$ and $\kappa_\mu$.

The \Sch equation for the Hamiltonian $\tilde H_{\rm eff}$ and the wave function  $\ket{\Psi}=\sum_{i\nu}p^{(i)}_\nu\ket{\xi^{(i)}_\nu}+\sum_\mu e_\mu\ket{\chi_\mu}$ leads to

\begin{equation}\label{Schroedinger-5}
\begin{split}
&\left(E-E_\nu^{(p,i)}\right) p^{(i)}_\nu = \frac{G X_\nu^{(i,\beta)}}{N}\sum\limits_\mu \Lambda_{\nu\mu}e_\mu,\\
&\left(E-E^{(ex)}_{\mu}\right) e_\mu =  \frac{G }{N}\sum\limits_{i = L,U} \sum\limits_{\nu} X_\nu^{(i,\beta)}\Lambda_{\nu\mu}p_{\nu}^{(i)}.
\end{split}
\end{equation}
We can exclude the exciton-exciton amplitudes $e_\mu$ from equations~\eqref{Schroedinger-5}; in the absence of hopping ($t\equiv0$) the resulting equation for $p_{i\nu}$ reduces to
\begin{equation}
\begin{split}
\left(E-E_{\nu}^{(p,i)}\right) p^{(i)}_\nu &= \frac{G^2 X_\nu^{(i,\beta)}}{2N (E-2E_0)}\\
&\times \sum\limits_{i' = L,U} \sum\limits_{\nu'} F_{\nu\nu'} X_{\nu'}^{(i',\beta)} p^{(i')}_{\nu'}
\end{split}
\end{equation}
with the kernel
\begin{equation}
F_{\nu\nu'}=N\left(\delta_{\nu,\nu'}+\delta_{\nu,-\nu'}\right) - \frac{2}{N}.
\end{equation}
The first term in the right-hand side of this equation describes the wave-vector-conserving scattering, while the second describes the formation of wave packets via scattering of non-interacting subsystem through the interacting one.

Within these notations, the amplitude for two photons being separated by $n$ lattice sites is
\begin{equation}
A(n)=\bra{\alpha_n}\Psi\rangle=\frac{1}{\sqrt{N}} \sum\limits_{i=L,U} \sum\limits_{\nu} p^{(i)}_\nu X_\nu^{(i,\alpha)} e^{-\frac{2\pi i\nu n}{N}}
\end{equation}
so that $A(0)=\sum_{i\nu}p^{(i)}_\nu X_\nu^{(i,\alpha)}/\sqrt{N}$ results from a collective effect of $p$-amplitudes that add up with a vanishing phase; large separation amplitudes are instead averaged out by the oscillating exponentials. The wider is the distribution of $p^{(i)}_\nu$, the larger $A(0)$ is expected. Using equations~(\ref{Schroedinger-5}) we get
\begin{equation}
\label{bun-2}
A(0)=\frac{G}{N\sqrt{N}} \sum\limits_{i = L,U} \sum\limits_{\nu} \frac{X_\nu^{(i,\alpha)} X_\nu^{(i,\beta)}}{\left( E-E_\nu^{(p,i)} \right)} \sum\limits_{\mu}\Lambda_{\nu\mu} e_{\mu}.
\end{equation}

Due to the mismatch between quantum numbers $\nu$ and $\mu$ the denominator $( E-E_\nu^{(p,i)} )$ is not a real pole. However, it plays an important role in the establishing of the bunching, which occurs when $E = E_\rho$ is resonant with the band of non-interacting states $E_\nu^{(p,i)}$ \eqref{AB-polaritons}; when $E_\rho < \min\left\{ E_\nu^{(p,i)} \right\} = E_{\nu=0}^{(p,i=L)}$ the two-photon wave function looks unperturbed and exhibits plane-wave-like oscillations. This criterium can be used as a good rule of the thumb when deciding on whether a state with a given energy shows bunching or not. It looks like as if excitons talked to each other via virtual excitations -- the eigenstates of the non-interacting subsystem. Indeed, the ``real" elementary excitations are one-polariton states, while the energies \eqref{AB-polaritons} do not have an independent physical meaning, except as a virtual scattering channel through which excitons interact.

Using the equality
\begin{equation}
C(s) = \bra{\gamma_s}\Psi \rangle = \sum\limits_\mu e_\mu g_s(\mu)
\end{equation}
following from $\ket{\Psi}$ representation via $A,B,C$- and $p,e$-amplitudes, we find that $2e_\mu = \sum_s g_s(\mu)C(s)$. For higher $\rho$ showing bunching, we can approximate $C$-amplitudes by bare exciton-exciton amplitudes \eqref{Cex(n)} times a normalization coefficient $X^{(\gamma)}_\rho$, which accounts for the presence of finite exciton-photon and photon-photon excitation in the total wave function of $\rho$-th eigenstate. We then obtain, using orthogonality of $g$-functions,
\begin{equation}
\begin{split}
\label{bun-4}
A_\rho(0) &\approx \frac{G X^{(\gamma)}_\rho}{2N^{\frac{3}{2}}} \sum\limits_{i = L,U} \sum\limits_{\nu} \frac{X_\nu^{(i,\alpha)} X_\nu^{(i,\beta)}}{(E_\rho - E_\nu^{(p,i)})}\\
&\times \left[\Lambda_{\nu \left(\mu=\rho\right)}+\Lambda_{\nu \left(\mu=-\rho\right)}\right].
\end{split}
\end{equation}

Equation~\eqref{bun-4} shows that only those state contribute into $A(0)$, which {\it simultaneously} have non-vanishing photonic and excitonic amplitudes, which is true only for the strong coupling region. Therefore, the larger part of the Brilloiun zone the strong coupling region occupies, the stronger is the effect of bunching. This explains the increase of the bunching efficiency with the increase of the lattice constant $a$: For larger $a$ the Brillouin zone of exciton is matched to a flatter part of the photon dispersion, and the strong coupling region is larger. This scale argument explains why in natural solids the kinematic interaction is a negligible effect, while in atomic systems it may lead to a qualitatively different behaviour.

\end{document}